\centerline{\bf Naked singularities and admissibility of initial conditions
}
\bigskip
\centerline{H. M.  Antia}
\centerline{Tata Institute of Fundamental Research, Homi Bhabha Road, Mumbai
400 005, India
}
\centerline{email: antia@tifrvax.tifr.res.in}
\vskip 2 cm
{\narrower

\centerline{\bf Abstract}

Gravitational collapse of a spherically symmetric  cloud has been
extensively studied to investigate the nature of resulting singularity.
However, there has been considerable debate about the admissibility of
certain initial density distributions. Using the Newtonian limit of the
equations governing collapse of a fluid with an equation of state
$p=p(\rho)$ it is
shown that the density distribution has to be even function of
$r$ in a spherically symmetric situation provided $dp/d\rho\ne0$.
Implications of this result on formation of strong naked singularities are
examined.
\bigskip
}
Gravitational collapse of spherically symmetric dust cloud has been
extensively studied [1-8] to understand the nature of resulting singularity.
Using an initial density distribution which is an even function of radial
distance, $r$  Newman [3] has shown that the singularity
is naked, but weak, when it just forms.
By dropping the assumption about evenness
of density it turns out that with suitable choice of initial density and
velocity the naked singularity could be strong [4,5]. However,
Unnikrishnan [6,7] has argued that such a density distribution is not
physical when a realistic equation of state with pressure is used.
He claims that density must be an even function of $r$.
On the other hand, using the special case of polytropic equation of state
in the Newtonian limit, Jhingan et al.~[8] have attempted to show that
density is not restricted to be an even function of $r$. Unfortunately, they
have not included the continuity equation  in their analysis and hence
their conclusions are not correct.
In this work, we show that using equations of fluid dynamics in the
Newtonian limit with an equation of state of the form $p=p(\rho)$,
(where $p$ is the pressure and $\rho$ the density) density has to
be an even function of $r$ in a spherically symmetric situation
provided $dp/d\rho\ne0$.

We consider spherically symmetric collapse of fluid with equation
of state $p=p(\rho)$, which is assumed to be continuous and
differentiable to any required order.
The usual equations of fluid dynamics in the Newtonian limit, i.e.,
the continuity equation and the momentum equation in the Eulerian
form are:
$$\eqalignno{
&{\partial \rho\over\partial t}+{1\over r^2}{\partial \over \partial r}
(r^2\rho v)=0 &(1)\cr
&\rho{\partial v\over\partial t}+\rho v{\partial v\over\partial r}=
-\rho {Gm\over r^2}-{\partial p\over\partial r} &(2)\cr}
$$
where $v$ is the radial component of velocity, $G$ is the gravitational
constant and 
$$
m=\int_0^r4\pi r^2\rho\;dr\eqno(3)
$$
is the total mass contained within a sphere of radius $r$.
It may be noted
that in [8] only the second equation was considered in the Lagrangian form
which is equivalent to Eq.2 of [8], while the continuity equation, which
is the Eq.3 in [8] was neglected.
Clearly, the continuity equation is automatically satisfied when considering
equilibrium situation ($v=0$), but while generalizing the
Lane-Emden equation to calculate collapse solutions, the continuity
equation will need to be added to determine the additional unknown,
namely, velocity.

Considering
a situation before the singularity is formed, we assume that the solution
is regular near the center and expand $\rho$, $v$, $p$ and $m$ in
power series about $r=0$:
$$\eqalign{
\rho(t,r)&=\sum_{i=0}^\infty \rho_i(t) r^i\qquad (\rho_0>0)\cr
v(t,r)&=r\sum_{i=0}^\infty v_i(t) r^i\cr
p(t,r)&=\sum_{i=0}^\infty p_i(t) r^i \cr
m(t,r)&=r^3\sum_{i=0}^\infty m_i(t) r^i \cr}\eqno(4)
$$
Here, $m$ and $p$ are not independent functions but are completely determined
by $\rho$. Thus we have
$m_i=4\pi\rho_i/(3+i)$ and $p_1=(dp/d\rho)\rho_1$, etc.
There is no constant term in expansion for velocity as $v$ must vanish
at the center in a spherically symmetric system.

Substituting these expansions in Eqs.~(1,2) and collecting all terms
with same powers of $r$, we can get the required equations to determine
the coefficients of expansion. Thus considering the terms of $r^0$ in
Eq.~(2) we get
$$
p_1={dp\over d\rho}\rho_1=0\eqno(5)
$$
Hence, if the equation of state is such that $dp/d\rho\ne0$, then
$\rho_1=0$ and $m_1=0$. This condition in physical terms simply states
that the pressure gradient which is the force, must vanish at the center
in a spherically symmetric situation. Now, considering terms of $r^1$
in Eq.~(1) and using the fact that $\rho_1=0$, we get $v_1=0$.
Similarly, considering terms of $r^2$ in Eq.~(2) and using $\rho_1=0$, $m_1=0$
and $v_1=0$, we get $p_3=(dp/d\rho)\rho_3=0$, and hence $\rho_3=0$ and
$m_3=0$. Again with all these conditions the terms of $r^3$ in
Eq.~(1) will give $v_3=0$. Thus the cubic terms in all expansions also
vanish. Continuing this process it is clear that all the odd terms in
the expansions (4) will vanish.

Thus it is clear that if $dp/d\rho\ne0$, the density is an even function
of $r$, while the radial component of velocity is an odd function of $r$.
Therefore, Unnikrishnan's [6,7] claim that solutions are analytic appears
to be correct, at least in the Newtonian limit. These results are likely
to be valid even for relativistic equations. However, it is clear that
these restrictions do not apply for the special case of dust ($p\equiv0$)
considered in most calculations of gravitational collapse.
 Nevertheless, if we think of dust as a
limiting case of fluid where the pressure gradient is small as compared to the
gravitational force, then the density should be considered as an even
function [7]. 

If we restrict to admissible solutions with $\rho$ and $v/r$ as even
functions then it turns out that the resulting solution cannot yield
a strong  naked singularity [5].
Strong  naked singularity
can form only when $\rho$ or $v/r$ have some odd terms and as such
it appears that these do not arise out of physically admissible initial
conditions. Even if we restrict to pure dust with no pressure, it turns
out that the solutions leading to strong naked singularity
are unstable (or non-generic) [8,9] and thus physically inadmissible.
Although Antia [9] considered only the special case of marginally bound
system to demonstrate the instability of these solutions, the proof is
equally valid for a general situation [8]. In the general case, for stability
it is essential that the solution is stable with respect to small
perturbations in both density and velocity. Considering the general case
of gravitational collapse of spherically symmetric dust cloud,
Singh and Joshi [5] have shown that strong  naked singularity can
form when certain quantities $Q_1$ and $Q_2$ defined by them vanish.
These restriction clearly makes the solutions unstable (or non-generic)
since a general perturbation to initial conditions will not satisfy these
restrictions.
Thus even though for any given initial density distribution it may be possible
to choose initial velocity such that collapse leads to a strong curvature
naked singularity [10], all such solutions are non-generic and are not likely
to arise in a physically realistic situation.

In fact the instability of solutions with strong naked singularities is
clear from results of Singh and Joshi [5], as on p 567 they write
``Also, the assumption $f_2\ne0$
is a natural one, because if the cloud is not marginally bound, it
will require a fine tuning of the velocity profile with
the density profile to make $f_2$ zero''. 
In fact, this is precisely the concept of stability or genericity.
However, on the next page  they consider cases where $f_3=0$, $f_4=0$
etc. If $f_2$ cannot be zero then there is no reason to assume that $f_3$
or $f_4$
can be zero, because that too will require fine tuning. Hence, all these
cases, which includes all solutions leading to strong naked singularities
should not be considered at all. This would confirm the claim
of Unnikrishnan [6,7] that the naked singularities in physical circumstances can
only be weak.

It may be noted that in all these calculations the singularity
is naked only when it just forms. Irrespective of initial conditions,
gravitational collapse ultimately results in a formation of
event horizon and hence black holes appear to be the only stable end
product of such a collapse.

We would also like to point out that such non-generic solutions
leading to strong naked singularity are not likely to
be realized in numerical solution [e.g., 11] of equations of
gravitational collapse. Since numerical solutions are inherently
approximate, even if we start with initial values which
satisfy the required conditions for formation of strong  naked singularity,
because of errors introduced in numerical solutions these conditions
will be violated at subsequent time and the resulting solution will
ultimately lead to weak singularity which results from a
generic initial conditions.
Similarly, the example given in [8] to illustrate that
absence of the apparent horizon until singularity formation does not imply
that the singularity is naked,
is not particularly relevant for numerical computations. In generic
solutions which are the only ones that would be expected in numerical work
such behavior is not seen.

Finally, we would like to point out that even for the case of pure dust
(i.e., $p\equiv0$),
if the density is assumed to be an even function of $r$, then from
the continuity equation it is clear that the velocity will be an
odd function of $r$ (i.e., $v_{2n+1}=0$, for $n=0,1,2,\ldots$) and once
again it is not possible to find any solution with strong 
naked singularity. Thus the solutions with strong naked singularity
mentioned in [11] must have odd terms in density at a general time $t$
even though at $t=0$ the density may be an even function of $r$.
This can be seen by writing the coefficient of $r$ in Eq.(2):
$${d\rho_1\over dt}+4(\rho_0v_1+\rho_1v_0)=0.\eqno(6)$$
Thus if at $t=0$, $\rho_1=0$ but $v_1\ne0$, then $d\rho_1/dt\ne0$
and hence the density will pick up odd terms. Such solutions have
no significance, as the density is an even function of $r$ at $t=0$ but
does not remain so at any other time.
Thus the choice of time $t=0$ at which the collapse is supposed to start
has to be very special to yield such solutions. In general, it should not
matter which epoch is chosen as $t=0$ and hence, if there is any reason
for the density to be an even function of $r$ it should be so at all times
$t$, in which case velocity must be an odd function of $r$ and it
is not possible to find any solution with strong naked singularity.
\bigskip

{\bf Acknowledgments:} I would like to thank Professor Pankaj S. Joshi
for encouraging me to write-up this work.

\bigskip
{\parindent=0 pt\everypar{\hangindent=20 pt}

[1] D. M. Eardley and L. Smarr, Phys.\ Rev.\ {\bf D19} (1979) 2239.

[2] D. Christodoulou, Commun.\ Math.\ Phys.\ {\bf 93} (1984) 171.

[3] R. P. A. C. Newman, Class.\ Quantum Grav.\ {\bf 3} (1986) 527.

[4] P. S. Joshi and T. P. Singh, Phys.\ Rev.\ {\bf D51} (1995) 6778.

[5] T. P. Singh and P. S. Joshi, Class.\ Quantum Grav.\ {\bf 13}
(1996) 559.

[6] C. S. Unnikrishnan, Gen.\ Rel.\ Grav.\ {\bf 26} (1994) 655

[7] C. S. Unnikrishnan, Phys.\ Rev.\ {\bf D53} (1996) R580.

[8] S. Jhingan, P. S. Joshi and T. P. Singh, Class.\ Quantum Grav.\ 
{\bf 13} (1996) 3057.

[9] H. M. Antia, Phys.\ Rev.\ {\bf D53} (1996) 3472.

[10] I. H. Dwivedi and P. S. Joshi, gr-qc/9612023.

[11] S. L. Shapiro and S. A. Teukolsky, Phys.\ Rev.\ Lett.\ {\bf 66} (1991)
994.

\end